\documentclass{aa}
\usepackage{graphicx}
\begin{document}
   \title{Constraints on parameters of B-type pulsators from combined multicolour photometry
   and radial velocity data.\\
     I. $\beta$ Cephei stars}

   \author{J. Daszy\'nska-Daszkiewicz$^{1,2}$, W. A. Dziembowski$^{2,3}$, A. A.
   Pamyatnykh$^{2,4}$}

   \offprints{J. Daszy\'nska-Daszkiewicz, \email{daszynska@astro.uni.wroc.pl}}

   \institute{{1} Instytut Astronomiczny, Uniwersytet Wroc{\l}awski,
   ul. Kopernika 11, 51-622 Wroc{\l}aw, Poland\\
   {2} Copernicus Astronomical Center, Bartycka 18, 00-716 Warsaw, Poland\\
   {3} Warsaw University Observatory, Al. Ujazdowskie 4, 00-478 Warsaw,
Poland\\
   {4} Institute of Astronomy, Russian Academy of Sciences,
   Pyatnitskaya Str. 48, 109017 Moscow, Russia\\
    }
   \date{Received ...; accepted ...}
   \abstract{We analyze data on pulsation amplitudes and phases for two $\beta$ Cephei
   stars, $\delta$ Cet and $\nu$ Eri. Str\"omgren photometry and  radial velocity
   measurements are used simultaneously to obtain constraints on mean parameters of the stars
   and identification of the excited modes. The inference about the radial mode order
   and mean star parameters is based on comparison of certain complex parameter, $f$,
   determined from data, with its counterpart derived from linear nonadiabatic
   modelling of stellar oscillations.
   The theoretical $f$ values are very sensitive to the adopted
   opacity data. In our modelling we rely on the  data from OPAL and OP
   projects. Significant differences were found. New seismic
   models of $\nu$ Eri were constructed with both the OPAL and OP
   opacities.
\\
   \keywords{stars: $\beta$ Cephei variables --
             stars: individual: $\delta$ Ceti, $\nu$ Eridani --
             stars: oscillation --
             stars: atmospheres
          }
}

   \titlerunning{Constraints on parameters of $\beta$ Cep stars from photometry
                 and radial velocity}
   \authorrunning{Daszy\'nska-Daszkiewicz et al.}
   \maketitle

\section{Introduction}

In our earlier papers (Daszy\'nska-Daszkiewicz, Dziem\-bowski \&
Pamyatnykh 2003, Daszy\'nska-Daszkiewicz et al. 2004) we
introduced a new asteroseismic probe of stellar structure. The
probe is the complex parameter, $f$, describing the ratio of the
relative luminosity variation to the relative radial displacement
of the surface. Empirical $f$ values may be derived from pulsation
data upon adopting an appropriate stellar atmosphere model. The
corresponding theoretical values come from linear nonadiabatic
calculations of stellar pulsation. Comparison of the empirical and
theoretical $f$ yields a probe of subphotospheric layer.
Specifically, this is the layer where the mode period is
comparable to the thermal time scale. It means that the layer is
not too shallow but still located mostly above the frequency-forming
interior. Thus, it is poorly probed by the frequency data.

Our new diagnostic tool was first applied to a sample of $\delta$
Scuti variables (see references above). In that work we were mostly
interested in probing the efficiency of subphotospheric convection,
whose treatment is a notoriously difficult aspect of stellar interior
physics. Calculated $f$ values proved very sensitive to the
parametrization of the mixing-length formalism.

In B-stars, treatment of convective transport is not a significant
factor in the modelling uncertainty but it is still of interest to ask
how well the empirical and theoretical $f$ values agree. The
driving effect in  B-type pulsators, such as $\beta$ Cephei and
SPB stars, arises in the layer of the iron opacity bump. This
layer has also the dominant influence on the $f$ parameter. What
is essential for the pulsational instability is the location of
the layer and the iron abundance. Location depends primarily on
the effective temperature. If the overall metallicity parameter,
$Z$, describes the iron abundance in the opacity bump layer, then
the two crucial parameters, both for the instability and the $f$
value, are $T_{\rm eff}$ and $Z$. In principle, these two
parameters can be derived from spectroscopy. However, there are
considerable uncertainties in the derived values which affect
accuracy of asteroseismic sounding. We have seen this in the case
of the $\nu$ Eri star (Pamyatnykh, Handler \& Dziembowski 2004,
Ausseloos et al. 2004), where the constraints on the convective
overshooting from the core were found to be strongly dependent on
$T_{\rm eff}$. In addition, we cannot be sure that the iron
abundance in the driving layer is the same as in the photosphere.
In fact, Pamyatnykh et al. (2004), facing problems with excitation
of certain modes in $\nu$ Eri, suggested that iron may be
accumulated around the opacity bump. Such an effect would surely
leave an imprint on the value of $f$. Perhaps such an accumulation
could also explain the unexpected discovery of a large number of
B-type pulsators in Magellanic Clouds (Ko{\l}aczkowski et al.
2004).

It is always important to use all available observables to check
the accuracy of stellar modelling. A comparison of empirical and
theoretical $f$ values provides a stringent test of models of
the atmosphere, envelope and pulsation. In addition, the comparison
offers an opportunity to test microscopic input physics,
especially the opacity data.

Determination of $f$ is possible only in conjunction with
identification of the $\ell$ degree for the excited mode. The
values of $\ell$ are important for association of measured
frequencies with normal modes of stellar oscillations. There are
several methods of $\ell$ identifications based on photometry or
on spectroscopy, but they required a priori knowledge of $f$ and,
frequently, the answer is not unique. In this paper we present the
first application of our method of simultaneous determination of
$\ell$ and $f$ from combined multi-colour photometry and radial
velocity data to $\beta$ Cephei stars.

In the next section we recall our method for simultaneous
determination of the $f$ parameter and the $\ell$ degree of an
observed mode from multicolour photometry and radial velocity
data. Section 3 is devoted to the analysis and interpretation of
observational data on $\delta$ Ceti star, which is a monoperiodic
$\beta$ Cep variable. The analysis of a multiperiodic $\beta$ Cep
star $\nu$ Eridani is a subject of Section 4. A summary and
discussion are given in Section 5.

\section{The method}

We repeat here the basic formulae underlying our method. The local
radial displacement of the surface element is written in the
standard form,
$$\delta r(R,\theta,\varphi)= R {\rm Re}\{ \varepsilon
Y_\ell^m {\rm e}^{-{\rm i}\omega t}\}, $$
where $\varepsilon$ is a small complex parameter fixing mode
amplitude and phase, whereas
$$Y^m_\ell(\theta,\phi)=(-1)^{\frac{m+|m|}{2}}
\sqrt{ \frac{(2\ell+1)(\ell-|m|)!}{(\ell+|m|)!} }
P_{\ell}^{|m|}(\cos\theta){\rm e}^{{\rm i} m\phi}.$$
With the adopted, non-standard, normalization of spherical harmonics, $|\varepsilon|$
is the r.m.s. value of $\delta r/R$ over the star surface. The corresponding changes of
the bolometric flux, ${\cal F}_{\rm bol}$, and the local gravity, $g$, are given
by
$$\frac{ \delta {\cal F}_{\rm bol} } { {\cal F}_{\rm bol} }= 
{\rm Re}\{ \varepsilon f Y_\ell^m {\rm e} ^{-{\rm i} \omega t} \},$$
and
$$\frac{\delta g}{g} = - \left( 2 + \frac{\omega^2 R^3}{G M}
\right) \frac{\delta r}{R}.$$
The complex parameter, $f$, which is the central quantity in this
work, has been already defined in the Introduction.

Since both $\varepsilon$ and $f$ may be regarded as constant in the
atmosphere, we can use the static plane-parallel approximation.
Then, the complex amplitude of the relative monochromatic flux
variation can be expressed as follows (see e.g.
Daszy\'nska-Daszkiewicz et al. 2002):
$$A^{\lambda}(i) = {\cal D}_{\ell}^{\lambda} ({\tilde\varepsilon}
f) +{\cal E}_{\ell}^{\lambda} {\tilde\varepsilon}, \eqno(1)$$
where
$${\tilde\varepsilon}\equiv \varepsilon Y^m_{\ell}(i,0),$$
%


$${\cal D}_{\ell}^{\lambda} = b_{\ell}^{\lambda} \frac14
\frac{\partial \log ( {\cal F}_\lambda |b_{\ell}^{\lambda}| ) }
{\partial\log T_{\rm{eff}}}, $$
$${\cal E}_{\ell}^{\lambda}= b_{\ell}^{\lambda} \left[ (2+\ell
)(1-\ell ) -\left( \frac{\omega^2 R^3}{G M}
 + 2 \right) \frac{\partial \log ( {\cal F}_\lambda
|b_{\ell}^{\lambda}| ) }{\partial\log g} \right],$$
and
$$b_{\ell}^{\lambda}=\int_0^1 h_\lambda(\mu) \mu P_{\ell}(\mu)
d\mu.$$
The $\lambda$ index identifies the passband. The partial
derivatives of ${\cal F}_\lambda |b_{\ell}^{\lambda}|$ may be
calculated numerically from tabular data. In this paper we use
data from Kurucz's models (2004). For the limb darkening
coefficients we take Claret's fits (2000). To convert the
amplitudes to magnitudes, the right hand side of Eq.(1) must be
multiplied by the factor $(-1.086)$.

A set of of observational equations for complex unknowns
$({\tilde\varepsilon} f)$ and ${\tilde\varepsilon}$ is obtained
from Eqs.(1) written for a number of passbands, $\lambda$ . On the
left-hand side we have measured amplitudes, $A^\lambda$, expressed
in the complex form. If we have data on spectral line variations,
the set of equations (1) may be supplemented with an expression
relating $\tilde\varepsilon$ to complex amplitudes of the first
moments, ${\cal M}_1^{\lambda}$,
$${\cal M}_1^{\lambda}= {\rm i}\omega R \left( u_{\ell}^{\lambda}
+ \frac{GM v_{\ell}^{\lambda} }{R^3\omega^2} \right)
\tilde\varepsilon. \eqno(2)$$
where
$$u_{\ell}^{\lambda} = \int_0^1 h_\lambda(\mu) \mu^2 P_{\ell}(\mu)
d\mu.$$
$$v_{\ell}^{\lambda} = \ell \int_0^1 h_\lambda(\mu) \mu \left(
P_{\ell-1}(\mu) - \mu P_{\ell}(\mu) \right)d\mu.$$

Another useful quantity from spectroscopy is the complex amplitude 
of equivalent width variations. This yields an additional constraint 
on $f$ and $\ell$. We plan to implement this in future work.

The set of Eqs. (1), possibly supplemented with Eq.(2), is solved
by the $\chi^2$ method for unknown parameters $\tilde\varepsilon$
and $(\tilde\varepsilon f)$.

Care must be taken in calculating complex observational
amplitudes. If the time dependence of the observed quantity, $O$,
is given in the form $A_O\sin(\omega t+\varphi)$ then the
corresponding complex amplitude is
$(A_O\sin\varphi,A_O\cos\varphi)$. Comparing the empirical and
theoretical $f$ values, we have to make sure that in the
nonadiabatic pulsation code the time dependence is ${\rm e}^{-{\rm
i}\omega t}$.

Effects of rotation are totally ignored in
equations (1) and (2). The approximation is essential because only
then unknown values of $i$ and $m$ may be absorbed into
$\tilde\varepsilon$. As long as the angular velocity of rotation
$\Omega$ is much lower than $\omega$ these equations provide a
good approximation. Effects of rotation enter the expression for
$A^{\lambda}$ through terms $\propto\Omega^2$ and
the expression for  ${\cal M}_1^{\lambda}$ through terms
$\propto\Omega$. Therefore use of Eq.(2) is less safe but, as we
will see later, it is essential for mode
identification in B-type stars. There is a need for improvement.
Fortunately the two stars considered in this paper are slow
rotators.

In our applications we take $\omega$ from observations, the
stellar radius, $R$, from the measured mean luminosity, $L$, and
effective temperature, $T_{\rm eff}$, while the mass, $M$, is
taken from evolutionary tracks.

\section{The monoperiodic $\beta$ Cep star $\delta$ Ceti}

\subsection{Data}

We derive mean parameters of the star relying on Str\"omgren
photometry from the catalogue of Stankov \& Handler (2005) and on
the Hipparcos parallax taking into account the Lutz-Kelker
correction (Lutz \& Kelker 1973). In Fig.\,1 we show position of
the star with its error box in the HR diagram. Table 1 presents 5
models which were chosen for analysis; all of them are consistent
with the observations. The models correspond to the central value
(D1) and to four values close to the edges of the error box
(D2-D5), and were fixed at $Z=0.02$.
\begin{figure}
\centering
\includegraphics[width=88mm,clip]{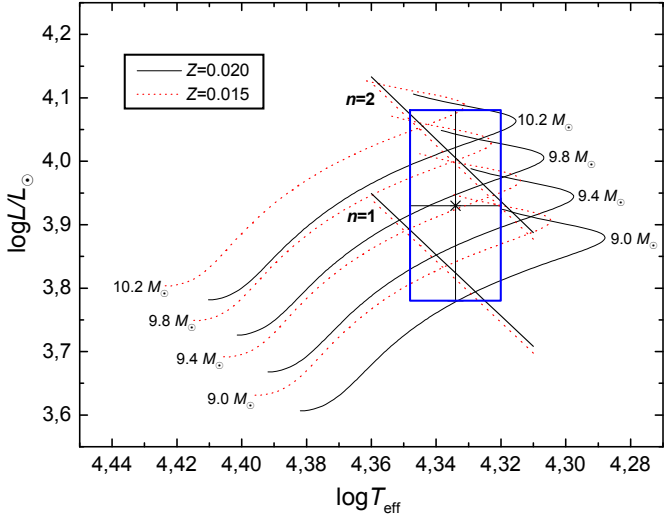}
\caption{ The observational error box for $\delta$ Ceti
in the HR diagram. The evolutionary tracks were calculated
for two values of the metallicity parameter $Z$.
The lines of constant fundamental $(n=1)$ and the first
overtone $(n=2)$ period equal to 0.161137 d. are shown.
These lines will be discussed later in this section.
}
\label{aaaaaa}
\end{figure} %
\begin{table}
\centering \caption{Parameters of selected models of $\delta$ Ceti.}
\begin{tabular}{ccccc}
\hline
& $M/M_\odot$ & $\log T_{\rm eff}$ & $\log L$ & $\log g$ \\
\hline
D1 &  9.70 &  4.3340 &  3.930 &  3.783 \\
D2 &  9.00 &  4.3229 &  3.806 &  3.830 \\
D3 &  9.20 &  4.3459 &  3.790 &  3.947 \\
D4 & 10.20 &  4.3232 &  4.041 &  3.650 \\
D5 & 10.40 &  4.3472 &  4.032 &  3.763 \\
\hline
\end{tabular}
\end{table}

We use here the amplitudes and phases in four Str\"omgren
passbands from Kubiak \& Seggewiss (1990). These photometric
observations were done in 1986. We derived the corresponding
radial velocity data from spectra taken in 1987 (Aerts et al. 1992).
The data used in our analysis are given in Table 2.
\begin{table}
\centering
\caption{Amplitudes and phases in radial velocity and in four
         Str\"omgren passbands for monoperiodic $\beta$ Cep star
         $\delta$ Cet with pulsation frequency 6.205875 c/d. }
\begin{tabular}{ccccccc}
\hline
     & A [km/s],[mag] & $\varphi$ [rad] \\
\hline
$V_{\rm rad}$ &  7.4323(723) &  1.6416(78)\\
$u$ &   0.0263(9) &  2.7758(156)\\
$v$ &   0.0143(9) &  2.7096(156)\\
$b$ &   0.0137(9) &  2.7291(156)\\
$y$ &   0.0128(9) &  2.7057(156)\\
\hline
\end{tabular}
\end{table}

\subsection{The $\ell$ degree of the excited mode}

The only detected mode in this star has been already identified as
$\ell=0$ from spectroscopy by Aerts et al. (1992) and from
photometry by Cugier et al. (1994). This identification was
confirmed by Cugier \& Nowak (1997), who included observations in
the UV range from IUE. However, in the photometric identification,
the theoretical $f$ values were adopted. Naturally, extracting
both $\ell$ and $f$ from observations sets higher requirements on
the data. Photometric data are insufficient.

\begin{table}
\centering
\caption{Values of $\tilde\varepsilon$, $f$ and $\chi^2$ for models of $\delta$ Ceti}
\begin{tabular}{cccccc}
\hline
& $\tilde\varepsilon_{\rm R}$ & $\tilde\varepsilon_{\rm I}$ &
$f_{\rm R}$ & $f_{\rm I}$ & $\chi^2$ \\
\hline
\multicolumn{6}{c}{$\nu$ = 6.20587 c/d, $\ell$ = 0} \\
\hline
 D1 & 0.00520(5) & 0.00036(4) & -9.70(18) & 2.56(12) & 1.03  \\
 D2 & 0.00570(5) & 0.00040(4) & -9.13(15) & 2.36(10) & 0.74  \\
 D3 & 0.00645(6) & 0.00045(5) & -8.37(14) & 2.07(10) & 0.90  \\
 D4 & 0.00435(5) & 0.00031(4) & -1.12(27) & 3.05(17) & 1.55  \\
 D5 & 0.00491(5) & 0.00035(4) & -1.01(22) & 2.69(14) & 1.28  \\
\hline
\end{tabular}
\end{table}
\begin{figure}
\centering
\includegraphics[width=88mm,clip]{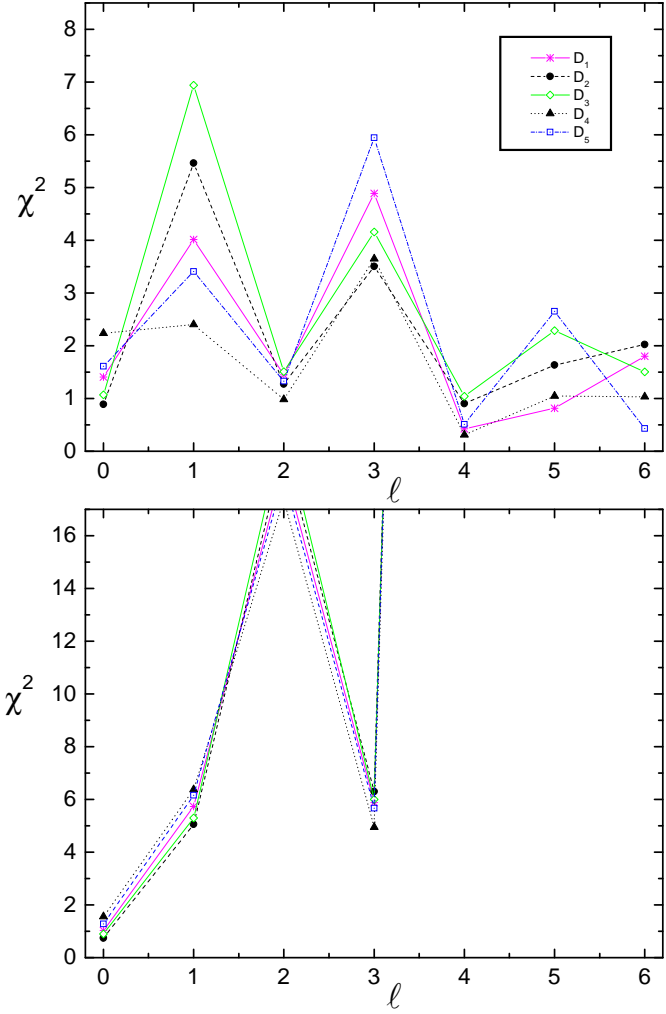}
\caption{
The $\chi^2$ per degree of freedom as a function of $\ell$ obtained
from the fit of photometric amplitudes and phases without radial velocity
data (upper panel) and with radial velocity data (lower panel).
The five lines correspond to models from Table 1 which are
consistent with the observations.
}
\label{aaaaaa}
\end{figure} %

It is clear from comparison of the $\chi^2(\ell)$ dependence,
shown in the two panels of Fig.\,2, that radial velocity data are
essential for determination of the spherical harmonic degree,
$\ell$. We see in the upper plot that without radial velocity
data, the discrimination between various even degrees is not
possible. The lower plot shows that the mode degree must be
$\ell=0$. In addition, the lower luminosity
models lead to a lower $\chi^2$. In Table 3 we provide our
solutions for the $\ell=0$ identification. With this
identification $\tilde\varepsilon$ is equal to the $\delta R/R$
amplitude.

\subsection{Discrimination between two possible radial modes}

Having determined the degree of the excited mode we still have to
consider its possible radial orders. In Fig.\,1 we show the
theoretical HR diagram with lines corresponding to a period of
0.16114 d for two different radial mode identifications. Both
fundamental and first overtone modes are acceptable. Now, for each
value of $Z$ and $n$ we have a one dimensional family of models to
consider.

We first check whether the $\chi^2$ behaviour may help us to
discriminate between the two possible choices of $n$. The result
shown in Fig.\,3 is rather negative. The difference between the two
choices is small, especially in the low temperature range, where
we reach the lowest $\chi^2$ regardless of the $n$ values.
On the same plot we show that the metallicity, [m/H],
microturbulence, $v_{\rm t}$, and opacity in the atmosphere
models do not matter very much for the $\chi^2$ values.
The metallicity affects the analysis in two ways.
The first is through the evolutionary models, hence the surface
gravity, the second is through the [m/H] parameter in the
atmosphere.
\begin{figure}
\centering
\includegraphics[width=88mm,clip]{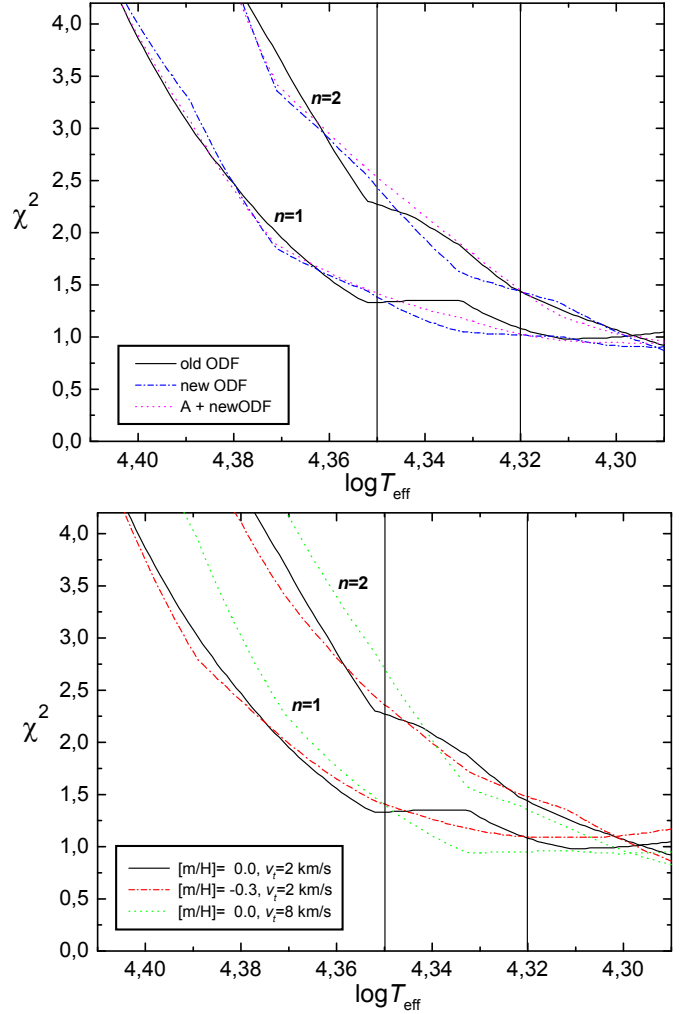}
\caption{ The $\chi^2$ per degree of freedom as a function of
$\log T_{\rm eff}$  for two possible identifications of the radial
mode order. The vertical lines show the allowed $T_{\rm eff}$ range,
consistent with the error box shown in Fig.\,1. In the upper panel
we show the effect of use of the new opacity distribution function
(ODF) in the atmosphere and enhancement in O, Ne, Mg, Si,  S, Ar,
Ca and Ti (A+newODF), following Kurucz (2004). In the lower panel
we compare the effect of different metallicity parameters, [m/H],
and microturbulence velocities, $v_{\rm t}$. } \label{aaaaaa}
\end{figure}

The lesson from the $\chi^2$ analysis is that the mode is definitely
the radial one and that there is some preference for lower
effective temperature and luminosity within the error box.
However, the discrimination between metallicities and two radial
mode orders was not possible. Much more may be learnt from
comparing the inferred nonadiabatic parameter, $f$, with model
calculations.

\begin{figure}
\centering
\includegraphics[width=88mm,clip]{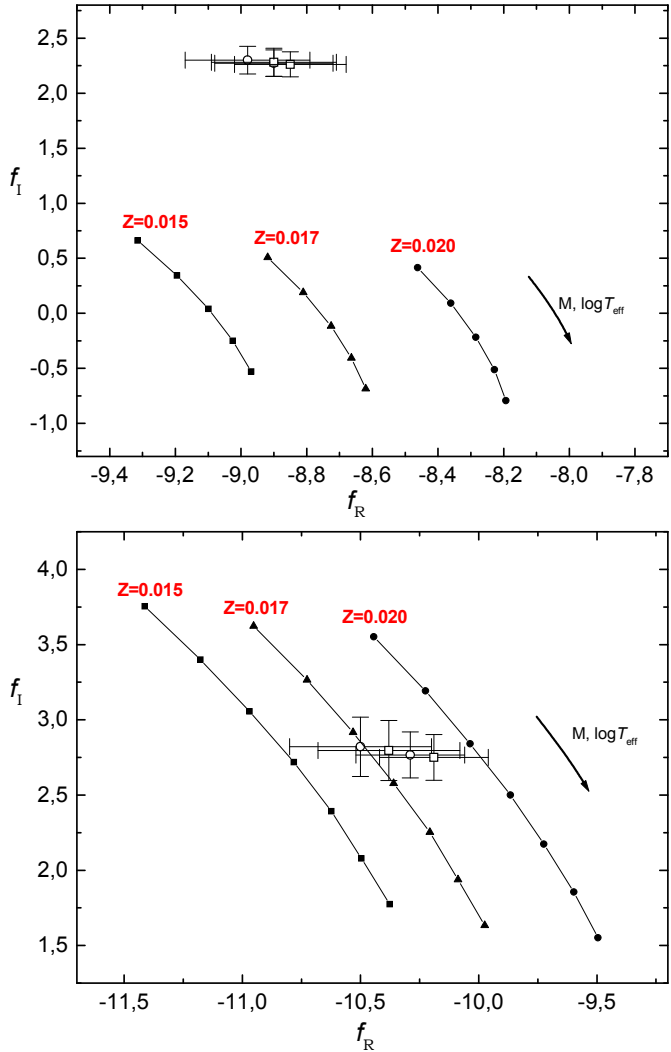}
\caption{ Comparison of the empirical and theoretical values of
$f$ assuming $Z$ and $T_{\rm eff}$ within the error box. The upper
and lower panels refer to the fundamental and first overtone
identification of the $\delta$ Cet mode, respectively.
The uncertainties in metallicity and effective temperature
are included both in the empirical and theoretical values of $f$.
The empirical values are taken from Table 3. The theoretical values
were calculated along lines of constant period (see Fig.\,1)
with the step in mass of $0.2M_\odot$.
In the upper panel, the masses at the top are 8.6, 8.8, 9.0 $M_\odot$ for
$Z$ = 0.015, 0.017, 0.020, respectively, whereas in the lower panel the masses
at the top are 9.2, 9.4, 9.6 $M_\odot$ for $Z$ = 0.015, 0.017, 0.020, respectively.
}
\label{aaaaaa}
\end{figure}
\begin{figure}[h]
\centering
\includegraphics[width=88mm,clip]{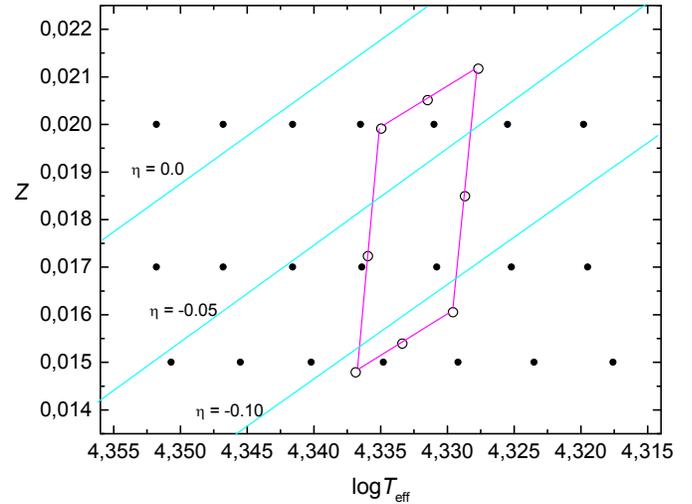}
\caption{ The ranges of $T_{\rm eff}$ and $Z$ allowed by
consistency of empirical and calculated values of $f$ assuming the
first overtone identification. The lines of constant instability
parameter, $\eta$, are shown. Negative values of $\eta$ correspond
to stability.} \label{aaaaaa}
\end{figure} %
%

\subsection{Inference from $f$-values and stability consideration}

As explained in the introduction, the calculated $f$ values must be
very sensitive to the iron abundance in the driving zone. The
effect is well seen in Fig.\,4. We also see the sensitivity to the
radial mode identification. Comparison of the upper and lower
panel strongly suggests that the excited mode is the first
overtone. This is the same identification as proposed by Cugier et
al. (1994) and supported by Cugier \& Nowak (1997). However there
is a problem with this identification, because in the allowed
ranges of $T_{\rm eff}$ and $Z$, the first overtone is stable, as
shown in Fig.\,5. The adopted measure of instability is $\eta$,
which varies between $-1$ and $+1$. For unstable modes we have
$\eta>0$. Though the mode is not far from the instability domain,
we regard the discrepancy as serious because in the same range of
parameters the fundamental mode is definitely unstable. This is a
typical property of $\beta$ Cep star models, in which the lower
frequency radial and neighbouring nonradial modes are
preferentially unstable.

We may think about various sources of the discrepancy in
empirical as well as in theoretical values of $f$. The effect
influencing the latter ones may be related to the opacity
data. Fortunately, we have two independent sources of these data.
Our results so far were obtained with the tabular opacities from the
OPAL project (Iglesias \& Rogers 1996). As the alternative we have
latest tables based on the Opacity Project (Seaton \& Badnell
2004, Badnell et al. 2005, Seaton 2005), hereafter OP.
As for the relative heavy element abundance, the OPAL data use the GN93
(Grevesse \& Noels 1993)  and the OP data
use S92 (Seaton et al. 1994) solar mix. The two mixes differ only slightly.
In particular, the S92 mix has a Fe abundance higher by 2.5\%.
Different opacities lead only to small differences in the
evolutionary tracks. The OP tracks are fainter by some 0.02 dex.
Surprisingly, a large difference was found in the $f$ values. Now,
there is no clear discrimination between the fundamental and first
overtone based on the comparison of the theoretical and empirical
$f$ values. Plots in Fig.\,6 show that both identifications are at
most marginally within with the error box. Note, however, that
while for the fundamental mode identification all acceptable
models lie well within the instability domain ($\eta>0$), the first
overtone identification places the models in the stability
domain. This is why we prefer the fundamental mode identification,
which implies, for instance, that luminosity is lower by 0.33 dex
and mass is lower by 2 $M_\odot$ than in the case of the first
overtone identification. Still, improvements are needed for a
convincing discrimination between the two possibilities.
\begin{figure}
\centering
\includegraphics[width=87mm,clip]{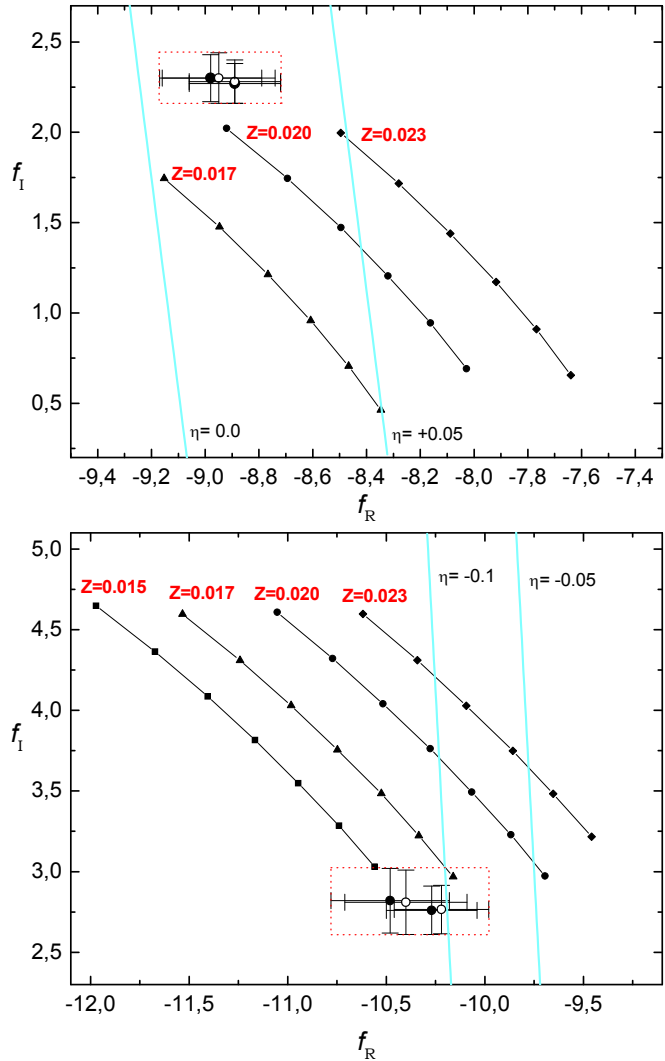}
\caption{ The same as Fig.\,4 but calculated with the new OP opacity
tables. In addition, lines of constant values of the instability
parameter, $\eta$, are shown.
In the upper panel, the masses at the top are 8.8, 8.8, 9.0 $M_\odot$
for $Z$ = 0.017, 0.020, 0.023, respectively, whereas in the lower panel
the masses at the top are 9.2, 9.4, 9.6, 9.8 $M_\odot$
for $Z$ = 0.015, 0.017, 0.020, 0.023, respectively.
The step in mass is $0.2M_\odot$.
}
\label{aaaaaa}
\end{figure} %

\section{The multiperiodic $\beta$ Cep star $\nu$ Eridani}

This star has been the object of the recent photometric (Handler et al. 2004)
and spectroscopic campaigns (Aerts et al. 2004, de Ridder et al. 2004).
The number of detected modes is much larger than in any other
star of this type. For several of them we have unambiguous mode
identification, which was supported by seismic model constructions
(Pamyatnykh, Handler \& Dziembowski 2004, Ausseloos et al. 2004).

Extracting $f$ values from data for a multiperiodic object is
clearly advantageous because we get independent constraints on
mean stellar parameters, such as $T_{\rm eff}$ or $Z$, from each
mode.

We adopted the mean parameters of $\nu$ Eri after Pamyatnykh,
Handler \& Dziembowski (2004), where details about derivation are
given. In Fig.\,7, we show the star position in the HR diagram
with the error box. Evolutionary tracks plotted in the figure were
calculated for two indicated $Z$ values, assuming no convective
overshooting. Seismic models of Pamyatnykh, Handler \& Dziembowski
(2004) were obtained at $Z=0.015$. Table 4 presents 5 models
which were chosen for analysis, all of them are consistent with
the observations. As in the case of $\delta$ Ceti, these five
models correspond to the central value (N1) and to four values
close to the edges of the error box (N2-N5), and were fixed at $Z=0.02$.
\begin{figure}
\centering
\includegraphics[width=88mm,clip]{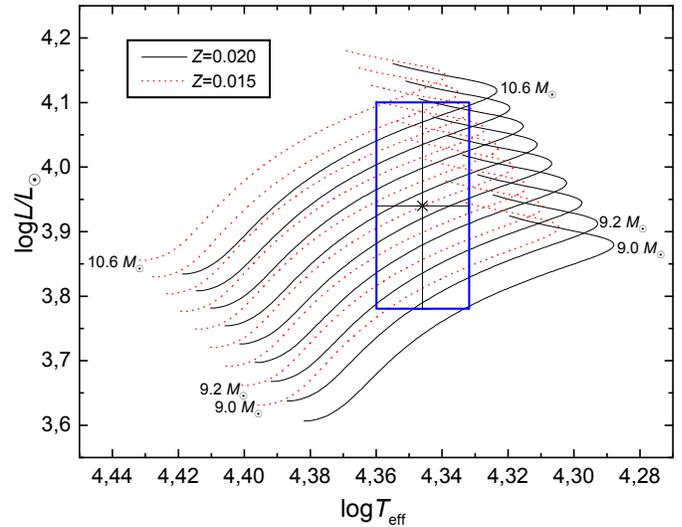}
\caption{ The observational error box for $\nu$ Eri on the HR
diagram. The evolutionary tracks for two values of the metallicity
parameter $Z$ are shown.}
\label{aaaaaa}
\end{figure}
\begin{table}
\centering \caption{Parameters of selected models of $\nu$
Eridani.}
\begin{tabular}{ccccc}
\hline
& $M/M_\odot$ & $\log T_{\rm eff}$ & $\log L$ & $\log g$ \\
\hline
N1 &  9.90 &  4.3460 &  3.940 &  3.830 \\
N2 &  9.20 &  4.3344 &  3.823 &  3.869 \\
N3 &  9.40 &  4.3589 &  3.792 &  4.007 \\
N4 & 10.40 &  4.3331 &  4.059 &  3.681 \\
N5 & 10.60 &  4.3584 &  4.044 &  3.805 \\
\hline
\end{tabular}
\end{table}
%

\subsection{Data}

Photometric amplitudes and phases used in our analysis of
this star were kindly provided to us by M. Jerzykiewicz, who
reanalyzed data from the multisite observations carried out
between October 2002 and February 2003 (Handler et al. 2004). To
determine the amplitudes and phases of the radial velocity, we
calculated the first moment of the line profile of SiIII 4553\AA~
using spectra obtained during the last spectroscopic campaign
(Aerts et al. 2004). We stress that it is important to use
contemporaneous data in this case because amplitudes and phases do
change at the level exceeding errors. It is also important that
the same frequencies are used in analysis of spectroscopic data.
The latter was the reason why we could not rely on results of De
Ridder et al. (2004), who analyzed data from the same campaigns.
At the moment, we have sufficiently accurate data only for the
four dominant peaks. We have amplitudes and phases in three
Str\"omgren passbands, $uvy$, and the radial velocity data.
The data used in our analysis are listed in Table 5.
\begin{table}
\centering
\caption{Amplitudes and phases in radial velocity and in four
    Str\"omgren passbands for four dominant frequencies in $\nu$ Eri.}
\begin{tabular}{ccccccc}
\hline
     & $A$ [km/s],[mag] & $\varphi$ [rad]  \\
\hline
\multicolumn{3}{c}{$\nu_1= 5.76326$ c/d} \\
\hline
$V_{\rm rad}$ & 22.017(116) &  5.473(5)\\
$u$ &   0.07345(20) &  0.616(3)\\
$v$ &   0.04103(14) &  0.590(3)\\
$y$ &   0.03685(13) &  0.579(4)\\
\hline
\multicolumn{3}{c}{$\nu_2=  5.65391$ c/d} \\
\hline
$V_{\rm rad}$ &      9.049(115) &  0.419(13) \\
$u$ &   0.03793(21) &  1.982(5)\\
$v$ &   0.02645(14) &  1.982(6)\\
$y$ &   0.02506(13) &  1.975(5)\\
\hline
\multicolumn{3}{c}{$\nu_3=  5.62009$ c/d} \\
\hline
$V_{\rm rad}$ &      8.272(112) &  0.129(14) \\
$u$ &   0.03464(21) & 1.681(6)\\
$v$ &   0.02388(15) & 1.691(6)\\
$y$ &   0.02267(14) & 1.684(6)\\
\hline
\multicolumn{3}{c}{$\nu_4=  5.63715$ c/d} \\
\hline
$V_{\rm rad}$ &      7.703(105) &  3.033(14) \\
$u$ &   0.03217(22) & 4.537(7)\\
$v$ &   0.02244(15) & 4.537(7)\\
$y$ &   0.02105(14) & 4.544(7)\\
\hline
\end{tabular}
\end{table}

\subsection{The $\ell$ and $f$ values}

Like in the case of $\delta$ Cet, only after combining photometric
and spectroscopic data, could we get a unique $\ell$
identification for the dominant modes in $\nu$ Eri. Our $\ell$
degrees are the same as those of our predecessors. However, unlike
them, we did not assume the $f$ values.
\begin{figure}[h]
\centering
\includegraphics[width=88mm,clip]{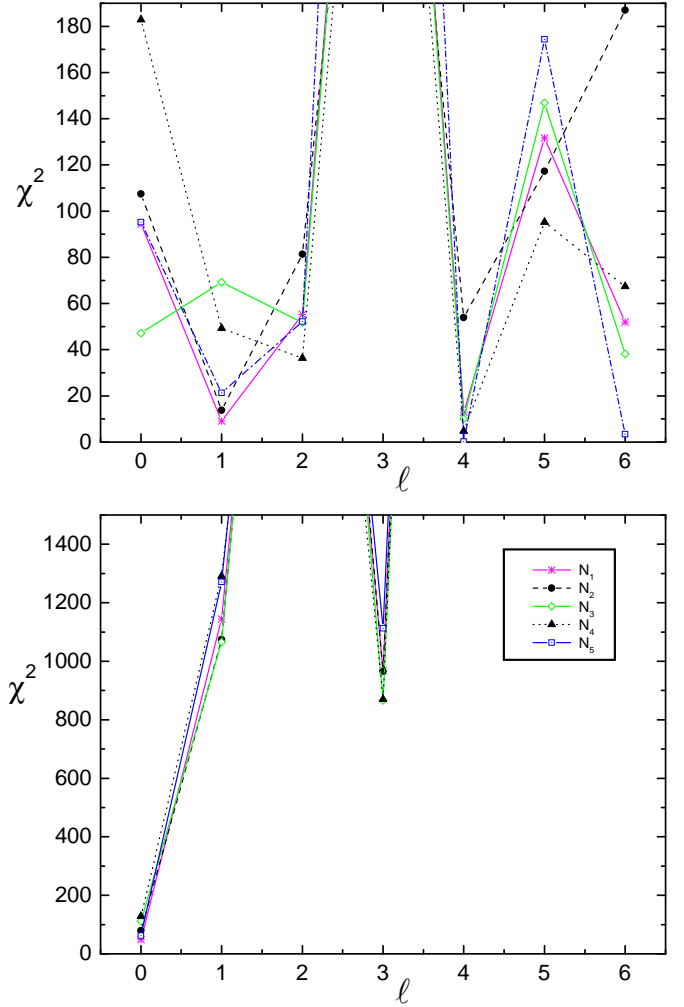}
\caption{The $\chi^2$ per degree of freedom as a function of
$\ell$ obtained from the fit of photometric amplitudes and phases
without radial velocity data (upper panel) and with radial
velocity data (lower panel), for the dominant frequency,
$\nu_1$. The five lines correspond to models from Table 1
which are consistent with the observations.
}
\label{aaaaaa}
\end{figure} %
\begin{figure*}
\centering
\includegraphics[width=17.5cm,clip]{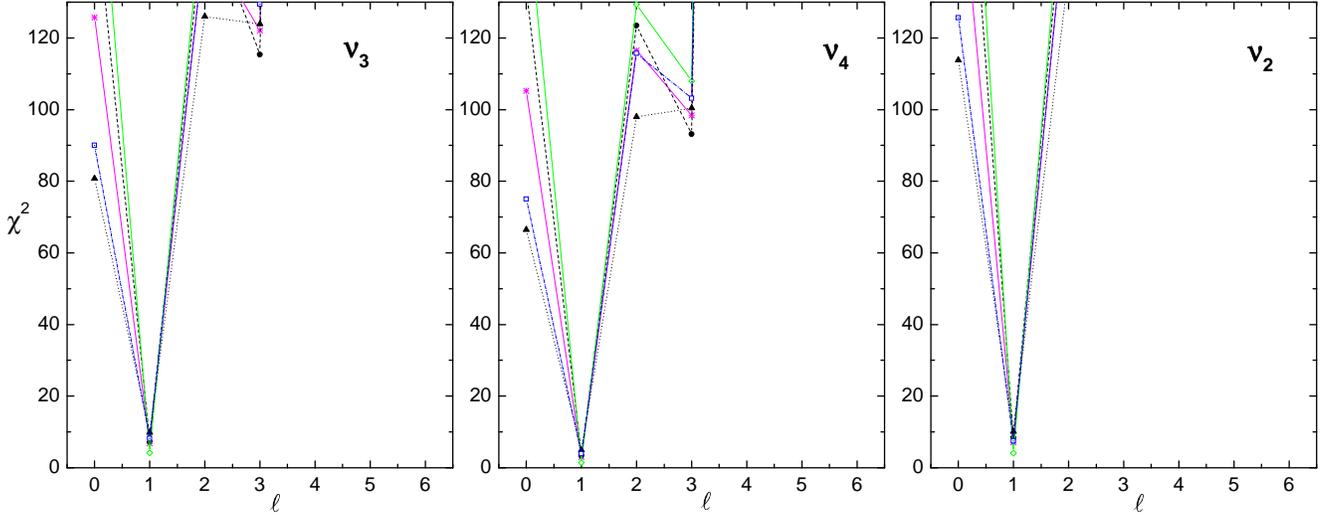}
\caption{ The same as in the lower plot of Fig.\,8 but for the
close frequency triplet, $\nu_3, \nu_4, \nu_2$ ($\ell=1, m=-1,0,+1$). } 
\label{aaaaaa}
\end{figure*} %
\begin{table}
\centering
\caption{Values of $\tilde\varepsilon$, $f$ and $\chi^2$ for models of $\nu$ Eridani}
\begin{tabular}{cccccc}
\hline
& $\tilde\varepsilon_{\rm R}$ & $\tilde\varepsilon_{\rm I}$ &
$f_{\rm R}$ & $f_{\rm I}$ & $\chi^2$ \\
\hline
\multicolumn{6}{c}{$\nu_1$ = 5.76326 c/d, $\ell$ = 0} \\
\hline
 N1 & -0.0125(6)  & -0.0118(6)  &  -9.2(4) & 0.8(4) &  48.4 \\
 N2 & -0.0133(8)  & -0.0126(8)  &  -8.8(5) & 0.7(5) &  79.6 \\
 N3 & -0.0152(11) & -0.0143(11) &  -8.1(5) & 0.6(5) & 112.0 \\
 N4 & -0.0104(8)  & -0.0099(8)  & -10.6(7) & 0.9(7) & 128.4 \\
 N5 & -0.0119(7)  & -0.0113(7)  &  -9.6(4) & 0.8(4) &  61.8 \\
\hline
\multicolumn{6}{c}{$\nu_3$ = 5.62009 c/d, $\ell$ = 1, $m$ = -1} \\
\hline
 N1 &  0.0010(3)  & -0.0080(3)  &  -9.0(3) & 0.1(3) &  6.81 \\
 N2 &  0.0011(3)  & -0.0085(3)  &  -8.4(3) & 0.1(3) &  7.27 \\
 N3 &  0.0012(3)  & -0.0094(3)  &  -7.5(2) & 0.1(2) &  4.15 \\
 N4 &  0.0009(3)  & -0.0067(3)  & -10.9(5) & 0.1(5) &  9.84 \\
 N5 &  0.0010(3)  & -0.0075(3)  &  -9.6(4) & 0.1(4) &  8.22 \\
\hline
\multicolumn{6}{c}{$\nu_4$ = 5.63715 c/d, $\ell$ = 1, $m$ = 0} \\
\hline
 N1 &  0.0008(2)  &  0.0074(2)  &  -9.0(2) & 0.5(2) &  3.02 \\
 N2 &  0.0009(2)  &  0.0079(2)  &  -8.4(2) & 0.5(2) &  3.40 \\
 N3 &  0.0010(1)  &  0.0088(1)  &  -7.5(1) & 0.5(1) &  1.46 \\
 N4 &  0.0007(2)  &  0.0063(2)  & -10.9(3) & 0.6(3) &  4.95 \\
 N5 &  0.0008(2)  &  0.0070(2)  &  -9.6(3) & 0.6(3) &  3.93 \\
\hline
\multicolumn{6}{c}{$\nu_2$ = 5.65391 c/d, $\ell$ = 1, $m$ = 1} \\
\hline
 N1 &  0.0036(3)  & -0.0080(3)  &  -9.1(3) & 0.1(3) &  7.07 \\
 N2 &  0.0038(3)  & -0.0086(3)  &  -8.5(3) & 0.1(3) &  8.34 \\
 N3 &  0.0042(3)  & -0.0094(3)  &  -7.5(2) & 0.1(2) &  4.06 \\
 N4 &  0.0030(3)  & -0.0068(3)  & -11.0(5) & 0.1(5) & 10.07 \\
 N5 &  0.003493)  & -0.0076(3)  &  -9.7(3) & 0.1(3) &  7.49 \\
\hline
\end{tabular}
\end{table}
In Fig.\,8, we compare $\chi^2$ values obtained with and
without the radial velocity data for the dominant mode, $\nu_1$,
which has been identified as $\ell=0$ (see e.g. De Ridder et al.
2004) on the base of photometric data alone but upon assuming
theoretical $f$. We see in the upper part of the figure
that our method without spectroscopic data does not
allow for $\ell$ determination. In the case of combined  data, the
minimum of $\chi^2$ is indeed very deep, however the value of
$\chi^2$ is large ($\approx 50$). Also one gets a very large $\chi^2$
($\approx40$) adopting $f$ from theory and relying only
on photometric data (De Ridder et al., 2004). For the three modes,
$\nu_2,\nu_3, \nu_4$, forming the close triplet, we confirm in
Fig.\,9 the $\ell=1$ identification. In Table 6 we provide
results of our analysis for all four modes but only $\ell$
leading to the $\chi^2$ minima. The minima for the triplet are
significantly lower than for the $\ell=0$ mode, but still larger
than 1. All these modes have close frequencies thus they should
have similar values of $f$. At frequencies around the fundamental
radial mode and above, these values are nearly $\ell$-independent,
as long as $\ell$ is not large. It is so because for such modes
pulsation is vertical in the layers contributing to $f$. 
Fig.\,10 shows that indeed all four $f$ values determined 
from the data are close, which is reassuring.
However, the large $\chi^2$ are of concern. There are two possible
explanations: either the model is inadequate or the errors 
are underestimated.

Let us begin with the first possibility. Neglect of the effect of rotation
in the model is fully justified because the
star is a very slowly rotating object. The ratio of rotation to
pulsation frequency is $\sim10^{-3}$. More suspect may seem our
use of linear relation between the displacement and the the
relative flux perturbation which is only an approximation, but it
seems justified. Consider the $\nu_1$ mode, which was identified
as radial and for which we determined the highest $\chi^2$ (still
by far the lowest than with other identification). With such
an identification we have $|\varepsilon|=|\tilde\varepsilon|=0.012$
 and $|\varepsilon f|=0.11$. The maximum relative
perturbation temperature is about 0.03, which is small. The most
natural interpretation of the higher order peaks (harmonics and
combinations) found e.g. by Handler et al. (2005) is a
nonlinearity of the $A^\lambda(\varepsilon)$ dependence. The
lowest-order nonlinear departure from Eq.(1) is cubic. There is a
number of third peaks but there is none at $3\nu_1$. The peak at
$2\nu_1$ is seen with the amplitude 1/16 of the $\nu_1$ peak. We
use these facts as an argument for the applicability of Eq.(1) 
in our case.

The next questionable part of our method concerns usage of
equilibrium atmosphere models for evaluation of the flux
derivatives in Eq.(1). Is the assumption that $\delta {\cal F}$
and $\delta r$ are constant within the atmosphere justified ? Are
the tabular data sufficiently accurate ? After the discussion of Cugier
et al. (1994), we are confident about a positive answer to the first
question. We are less confident in the case of the second question
but it seems unlikely that this is the cause of large $\chi^2$
because in the case of $\delta$ Ceti the $\chi^2$ minima were
about\,1. We used atmospheric models from the same source and the
stellar parameters of these two stars are very close.

Having rejected the inadequacy of the model, we turn to the error
estimate. Let us first note that the main contribution to $\chi^2$
(above 90\%) comes from photometric data and that, in the case of
the $\nu_1$ peak, we require a 7 times error increase of the
photometric measurement to get $\chi\approx1$. The errors given in
Table 6 could be underestimated. The differences in amplitudes
derived from various data sets exceeds the errors by a factor 5. The
problem was briefly discussed by Jerzykiewicz et al. (2005) who
suggest only a factor of 2 error enhancement. A more sophisticated
approach to the error analysis is needed but it is beyond the
scope of the present work. We proceed further relying on
determined $f$ values, being encouraged by the their consistency
and reasonably low errors as well by the clear $\ell$
discrimination.


%
\begin{figure}
\centering
\includegraphics[width=87mm,clip]{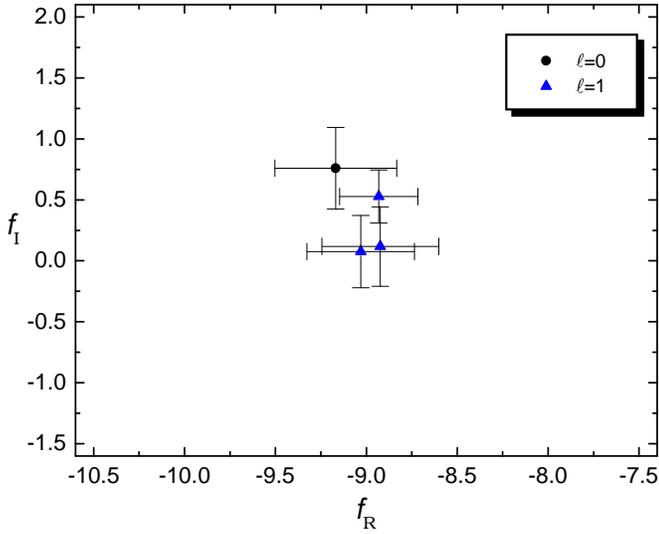}
\caption{ Empirical $f$ values for $\nu_1$, $(\ell=0)$, and the
close triplet $\nu_2, \nu_3, \nu_4$, $(\ell=1)$ for
$\log T_{\rm eff}=4.355$ and $\log L/L_{\odot}=3.969$.}
\label{aaaaaa}
\end{figure} %
In our comparison of empirical and theoretical values of $f$, we
will consider only the $\ell=0$ mode because the errors are the
lowest in that case. It is also safer than using the average from
the four modes  because there may be a noticeable effect of
difference in frequencies. Values of $f$ for modes at more distant
frequencies would be much more useful for estimating the $f(\nu)$
dependence, which is of interest. Unfortunately, we did not
succeed in a meaningful determination of $f$ for any high
frequency mode in $\nu$ Eri.

Comparison of the two panels of Fig.\,11 shows clearly that the
fundamental mode identification is preferred. The same
identification was obtained by Pamyatnykh, Handler \& Dziembowski (2004)
and Ausseloos et al. (2004) who constructed seismic
models $\nu$ Eri based on a simultaneous fit of the radial
and dipole mode frequencies measured in the star.

As in the case of the $\delta$ Cet model, we find that opacity
significantly influences the theoretical $f$ but, regardless of the
source of the opacity data, the fundamental radial mode is the only
acceptable identification. Comparing  Fig.\,12 with the upper panel
of Fig.\,11, we see somewhat better agreement between the empirical
and theoretical values  when the OP opacities are used. However,
the inference on the star parameter is significantly different. 
For instance, using OP data we get the mass higher by about
$0.4M_\odot$, $Z$ lower by about 0.0015, and $\log T_{\rm eff}$
higher by 0.01. The above conclusions are
based on stellar models constrained only by mean parameters from
photometry and the radial mode frequency, $\nu_1$. We will see
that our new seismic models lead us to somewhat different results.

\begin{figure}
\centering
\includegraphics[width=88mm,clip]{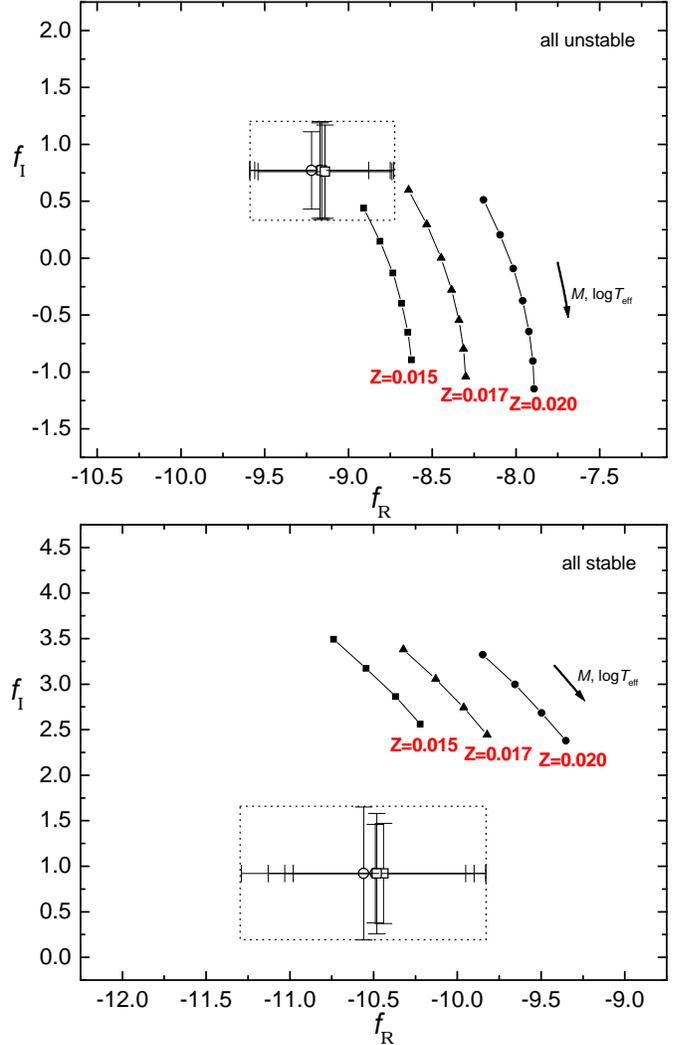}
\caption{ Theoretical and empirical $f$ for the radial mode
($\nu_1$) assuming various stellar parameters within the error
box. The upper panel assumes fundamental and the lower one the
first overtone mode identification.
The empirical values are taken from Table 6.
Theoretical $f$ were obtained adopting the three indicated values
of the metallicity parameter, $Z$, and the OPAL opacities.
In the upper panel, the masses at the top are 9.0, 9.0, 9.2 $M_\odot$
for $Z$ = 0.015, 0.017, 0.020, respectively, whereas in the lower panel
the masses at the top are 9.6, 9.8, 10.0 $M_\odot$
for $Z$ = 0.015, 0.017, 0.020, respectively.
The step in mass is $0.2M_\odot$.
}
\label{aaaaaa}
\end{figure} %
\begin{figure}
\centering
\includegraphics[width=88mm,clip]{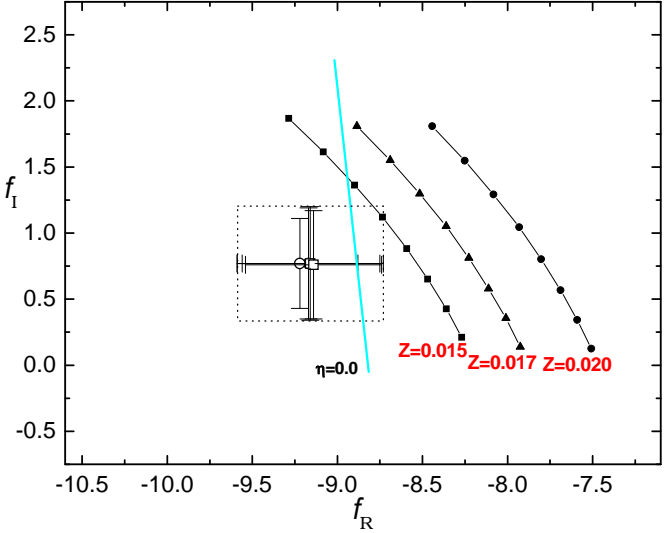}
\caption{ The same as in the upper panel of Fig.\,11 but with the
theoretical $f$ values calculated with the OP opacities. Models
lying to the right of the line marked with $\eta=0.0$ are unstable.
The masses at the top are 8.8, 9.0, 9.2 $M_\odot$
for $Z$ = 0.015, 0.017, 0.020, respectively. The step in mass is $0.2M_\odot$.
}
\label{aaaaaa}
\end{figure}
%

\subsection{New seismic models}

We constructed seismic models using both OPAL and OP data. Like
Pamyatnykh, Handler \& Dziembowski (2004), the construction uses
frequencies of the $\ell=0,p_1$, $\ell=1,g_1$ and $\ell=1,p_1$
modes. Models calculated with OPAL are slightly different than
those of Pamyatnykh, Handler \& Dziembowski (2004) due to a higher
accuracy demand on the frequency fit. The model parameters and the
$f$ values for $\ell=0,p_1$ are listed in Table 7. 
For each of the opacities, we
provide data for the models obtained with two values of the
overshooting parameter, $\alpha_{\rm ov}=0$ and 0.1. In the
seismic models, the influence of $\alpha_{\rm ov}$ on $f$ enters
indirectly through $T_{\rm eff}$ and $Z$. We may see that the
parameters of the models are significantly affected by the choice
of the opacity data. Models calculated with the OP data are cooler
by about 0.015 dex. The model calculated with $\alpha_{\rm
ov}=0.1$ is in fact somewhat outside the allowed range of $T_{\rm
eff}$. The OP models have higher $Z$ by about 0.003. We see
that largest differences between the OPAL and OP models are in the
imaginary part of $f$. Indeed the value of $f$ is a very sensitive
probe of the stellar opacities.

How the $f$ values from the four seismic models compare with
observations is shown in Fig.\,13. The corresponding empirical $f$
are also given in Table 8. For none of the opacity data is
the agreement fully satisfactory. 
The truth seems to be somewhere
in between, if a moderate overshooting is allowed. Relying only on
OPAL opacities, we may reach the agreement by increasing
$\alpha_{\rm ov}$ to about 0.2. However, a further increase would
leave the instability range.

\begin{table}
\centering
\caption{Seismic models of $\nu$ Eri calculated with the OPAL and
OP opacities. The $f$ values are given for $\ell=0,p_1$.}
\begin{tabular}{cccccccc}
\hline
 & $M/M_\odot$&$\alpha_{\rm ov}$&  $Z$& $\log T_{\rm eff}$& $\log L$&$f_R$&$f_I$\\
\hline
&&&OPAL&&&&\\
\hline
S1 & 9.808& 0.0&    0.0155& 4.3530&  3.959&  -8.56&  -0.63\\
S2 & 9.230& 0.1&    0.0145& 4.3424&  3.903&  -8.82&  -0.07\\
 \hline
&&&OP&&&&\\
\hline
S3 & 9.590& 0.0&    0.0185&  4.3396&  3.902&  -8.22&   1.20\\
S4 & 9.020& 0.1&  0.0175&  4.3284&  3.843&  -8.70&   1.70\\
\hline
\end{tabular}
\end{table}
\begin{table}
\centering
\caption{Values of $\tilde\varepsilon$, $f$ and $\chi^2$ for seismic
models of $\nu$ Eridani for the dominant frequency, $\nu_1$.}
\begin{tabular}{cccccc}
\hline
& $\tilde\varepsilon_{\rm R}$ & $\tilde\varepsilon_{\rm I}$ &
$f_{\rm R}$ & $f_{\rm I}$ & $\chi^2$ \\
\hline
\multicolumn{6}{c}{$\nu_1$ = 5.76326 c/d, $\ell$ = 0} \\
\hline
 S1 & -0.0126(6)  & -0.0120(6)  &  -9.2(3) & 0.8(3) &  40.2 \\
 S2 & -0.0128(7)  & -0.0121(7)  &  -9.1(4) & 0.8(4) &  55.8 \\
 S3 & -0.0126(7)  & -0.0120(7)  &  -9.2(4) & 0.8(4) &  56.4 \\
 S4 & -0.0127(7)  & -0.0121(7)  &  -9.2(4) & 0.8(4) &  61.9 \\
\hline
\end{tabular}
\end{table}
\begin{figure}
\centering
\includegraphics[width=88mm,clip]{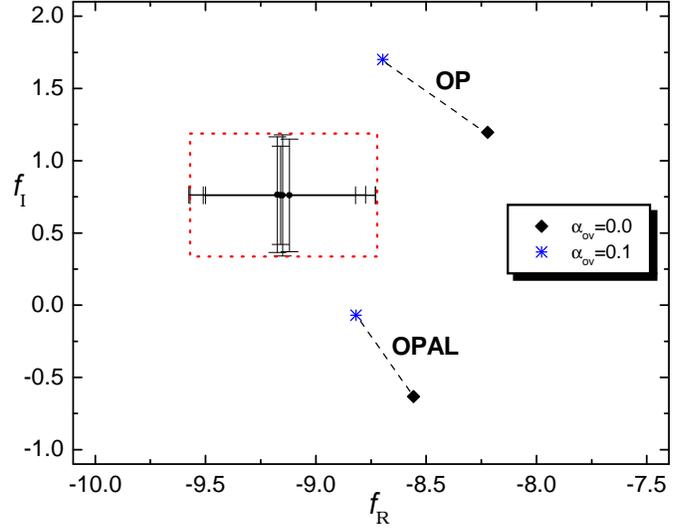}
\caption{ The f values for fundamental radial mode of the seismic
models constructed using opacity data from two sources (OPAL and
OP) and using two values of the overshooting parameter
($\alpha_{\rm ov}=0.0$ and 0.1). The empirical f values were
obtained assuming atmospheric parameters corresponding to the seismic models.
}
\label{aaaaaa}
\end{figure}

\section{Conclusions}

Using amplitudes and phases from simultaneous photometric and
spectroscopic observations of two $\beta$ Cep stars, we derived
the spherical harmonic degree, $\ell$, of excited modes and the
complex parameter, $f$, which may be directly compared with the
nonadiabatic theory of stellar oscillations. The $f$ parameter
determines bolometric flux amplitude at specified surface
displacement and it constitutes a new seismic probe of stellar
outer layers.

One of the stars is a monoperiodic pulsator $\delta$ Cet, the
other is a multiperiodic pulsator $\nu$ Eri. In the latter star,
we analyzed data on four dominant peaks. We found that unique
determination of mode degree  requires both photometric and
spectroscopic data. In our analysis we relied on the use of static
atmospheric models of Kurucz (2004).

Empirical values of $f$ were compared with the theoretical values
for relevant stellar models. The models employed the OPAL (Iglesias \& Rogers 1996)
and the OP tabular data (Seaton \& Badnell 2004, Badnell et al. 2005, Seaton 2005).
The evolutionary tracks are not much affected by the choice
of the data but significant differences are seen in the resulting $f$.

We confirmed the $\ell=0$ identification for the only mode
detected in $\delta$ Cet but, unlike our predecessors, we think
that the mode is more likely fundamental than the first overtone.
The reason is that the latter mode is stable over the allowed
range of mean star parameters. If indeed the mode is fundamental
then approximate consistency of the $f$ values is possible only
when the OP opacities are used.

Our identification of the four dominant peaks in the $\nu$ Eri
oscillation spectrum is the same as that of
Pamyatnykh, Handler \& Dziembowski (2004) and Ausseloos et al. (2004).
Consistency between the theoretical and empirical values of $f$
was achieved both with OPAL and OP data, but at significantly
different stellar parameters.

For none of the two opacity data did we achieve full agreement
between empirical $f$ and those derived from seismic models
fitting three dominant centroid frequencies in the oscillation
spectrum. As one can see, a perfect fit would be possible at some
intermediate opacities assuming a small overshooting parameter,
$\alpha_{\rm ov}$.

There are things to improve in our models of $\beta$ Cep stars.
As Pamyatnykh, Handler \& Dziembowski (2004)
pointed out, the standard models do not
account for excitation of modes at very low and very high
frequencies in $\nu$ Eri. We also cannot be fully satisfied with
our fit of empirical and theoretical $f$. These two problems may
be related. The solution to the first problem proposed by
Pamyatnykh, Handler \& Dziembowski (2004) was
a significant enhancement of the iron abundance in the driving
zone caused by the selective radiation pressure. Undoubtedly, this
will be reflected in the $f$ values. We plan to investigate the
effect when we have physically justified models of the enhancement.

Ideal seismic stellar models should account, within the errors,
not only for all measured frequencies but also for the $f$ values.
Such models will provide an ultimate test of our knowledge of the
stellar structure from the atmosphere to the deep interior. We are
not at this point yet. We believe that the most important result
of the present work is that our new seismic tool provides a
stringent probe of stellar opacities.


\begin{acknowledgements}
We are grateful to Mike Jerzykiewicz for kindly providing data on
photometric amplitudes and phases as well as for illuminating
discussion about possible sources of errors. We thank Mike Seaton
for detailed discussion of how to use his new opacity data, and
Conny Aerts for making spectra of $\delta$ Cet and $\nu$ Eri accessible.
We also thank the anonymous referee for stimulating comments.
JDD thanks the Foundation for Polish Science for supporting her
stay in the Copernicus Astronomical Center. 
The work was partially supported by Polish MNiI grant No. 1 P03D 021 28.
\end{acknowledgements}



\begin{thebibliography}{}

\bibitem[2004]{} Ausseloos, M., Scuflaire, R., Thoul, A., Aerts, C., 2004, MNRAS 355, 352

\bibitem[2004]{} Aerts, C., de Cat, P., Handler, G., et al., 2004, MNRAS 347, 463

\bibitem[1992]{} Aerts, C., de Pauw, M., Waelkens, C., 1992, A\&A 266, 294

\bibitem[2004]{} Badnell, N. R., Bautista, M. A., Butler, K., et al., 2005, MNRAS, in press (astro-ph/0410744)

\bibitem[2000]{} Claret, A., 2000, A\&A 363, 1081

\bibitem[1994]{} Cugier, H., Dziembowski, W. A., Pamyatnykh, A. A., 1994, A\&A 291, 143

\bibitem[1997]{} Cugier, H., Nowak, D., 1997, A\&A 326, 620

\bibitem[2002]{} Daszy\'nska-Daszkiewicz, J., Dziembowski, W. A.,
                 Pamyatnykh, A. A., Goupil, M-J. 2002, A\&A 392, 151

\bibitem[2003]{} Daszy\'nska-Daszkiewicz J., Dziembowski W. A., Pamyatnykh A. A., 2003, A\&A 407, 999

\bibitem[2004]{} Daszy\'nska-Daszkiewicz, J., Dziembowski, W. A., Pamyatnykh, A. A., Breger, M., Zima, W., 2004,
               Proc. IAU Symp. 224, eds. J. Zverko, W. W. Weiss, J. Ziznovsky and S. J. Adelman, p. 853

\bibitem[1993]{} Grevesse, N., Noels, A., 1993, in Origin and
Evolution of the Elements, eds. N. Pratzo, E. Vangioni-Flam and M.
Casse, Cambridge Univ. Press, p. 15

\bibitem[2004]{} Handler, G., Shobbrook, R. R., Jerzykiewicz, M. et al., 2004, MNRAS, 347, 454

\bibitem[1996]{} Iglesias, C. A., Rogers, F. J., 1996, ApJ 464, 943

\bibitem[2005]{} Jerzykiewicz, M., Handler, G., Shobbrook, R. R., et al., 2005, MNRAS, 360, 619

\bibitem[2004]{} Ko{\l}aczkowski, Z., Pigulski, A., Soszy\'nski, I., et al., 2004,
                 in Variable Stars in the Local Group, Proc. IAU Coll. 193,
           eds. D. W. Kurtz and K. R. Pollard, ASP Conf. Ser., Vol. 310, 225

\bibitem[1990]{} Kubiak, M., Seggewiss, W., 1990, Acta Astron. 40, 85

\bibitem[2004]{} Kurucz, R. L., 2004, http://kurucz.harvard.edu

\bibitem[1973]{} Lutz, T. E., Kelker, D. H., 1973, PASP 85, 573

\bibitem[2004]{} Pamyatnykh, A. A., Handler, G., Dziembowski, W. A., 2004, MNRAS 350, 1022

\bibitem[2004]{} de Ridder, J., Telting, J. H., Balona, L. A., et al., 2004, MNRAS 351, 324

\bibitem[2005]{} Seaton, M. J., 2005, MNRAS, in press, (astro-ph/0411010)

\bibitem[2004]{} Seaton, M. J., Badnell, N. R., 2004, MNRAS 354, 457

\bibitem[1994]{} Seaton, M. J., Yu Yan, Mihalas, D., Pradhan, A. K., 1994, MNRAS,
266, 805

\bibitem[2005]{} Stankov, A., Handler G., 2005, ApJS, 158, 193

\end{thebibliography}
\end{document}